\documentclass[aps,prl,twocolumn,showpacs,amsmath,amssymb,superscriptaddress,longbibliography]{revtex4-1}

\usepackage{graphicx}% Include figure files
\usepackage{dcolumn}% Align table columns on decimal point
\usepackage{bm}% bold math
\usepackage{fouriernc} % font style
\usepackage{color}
\usepackage{soul,color,xcolor}
\usepackage{epsfig}
\usepackage{fancyhdr}
\usepackage{epstopdf}
\include{graphic}

\begin{document}

\title{Meter-scale strong coupling between magnons and photons}

\author{Jinwei~Rao} \email{raojw@shanghaitech.edu.cn;}
\affiliation{School of Physical Science and Technology, ShanghaiTech University, Shanghai 201210, China}

\author{C. Y.~Wang}
\affiliation{School of Physical Science and Technology, ShanghaiTech University, Shanghai 201210, China}

\author{Bimu~Yao} \email{yaobimu@mail.sitp.ac.cn;}
\affiliation{State Key Laboratory of Infrared Physics, Shanghai Institute of Technical Physics, Chinese Academy of Sciences, Shanghai 200083, China}
\affiliation{School of Physical Science and Technology, ShanghaiTech University, Shanghai 201210,  China}

\author{Z. J.~Chen}
\affiliation{School of Physical Science and Technology, ShanghaiTech University, Shanghai 201210, China}

\author{K. X.~Zhao}
\affiliation{School of Physical Science and Technology, ShanghaiTech University, Shanghai 201210,  China}

\author{Wei~Lu} \email{luwei@shanghaitech.edu.cn;}
\affiliation{School of Physical Science and Technology, ShanghaiTech University, Shanghai 201210,  China}
\affiliation{State Key Laboratory of Infrared Physics, Shanghai Institute of Technical Physics, Chinese Academy of Sciences, Shanghai 200083, China}

\begin{abstract}

We experimentally realize a meter-scale strong coupling effect between magnons and photons at room temperature, with a coherent coupling of $\sim20$ m and a dissipative coupling of $\sim7.6$ m. To this end, we integrate a saturable gain into a microwave cavity and then couple this active cavity to a magnon mode via a long coaxial cable. The gain compensates for the cavity dissipation, but preserves the cavity radiation that mediates the indirect photon-magnon coupling. It thus enables the long-range strong photon-magnon coupling. With full access to traveling waves, we demonstrate a remote control of photon-magnon coupling by modulating the phase and amplitude of traveling waves, rather than reconfiguring subsystems themselves. Our method for realizing long-range strong coupling in cavity magnonics provides a general idea for other physical systems. Our experimental achievements may promote the construction of information networks based on cavity magnonics.

\end{abstract}

\maketitle
Many recent advances in quantum science and information technology have benefited from engineering strong, steerable interactions between systems with distinct properties \cite{pezze2018quantum,gross2017quantum,kimble2008quantum,tabuchi2015coherent,lachance2020entanglement}. In these hybrid systems, strong coupling between subsystems routinely forms in close proximity, for instance the formation of polaritons in quantum wells \cite{gibbs2011exciton}, strong photon-atom coupling in cavities \cite{mabuchi2002cavity}, optical mode interactions in microcavities \cite{scully1999quantum}. However, with the emerging trend of quantum network and long-range quantum entanglement, there is a strong expectation within the scientific community to overcome the limitations of near-field interaction and extend the coupling range beyond the typical wavelength \cite{richerme2014non,landig2016quantum,mottl2012roton,PRXQuantum2040314,wang2022giant, li2022coherent,gopalakrishnan2009emergent,maghrebi2017continuous,hofmann2012heralded}. The central quest is to find a way to bidirectionally transfer interactions between subsystems in the far field. For instance, when using a free-space laser beam, the strong coupling is achieved between a mechanical oscillator and atomic spins that are 1 meter apart \cite{karg2020light}. 

\begin{figure*} [htbp]
\begin{center}
\epsfig{file=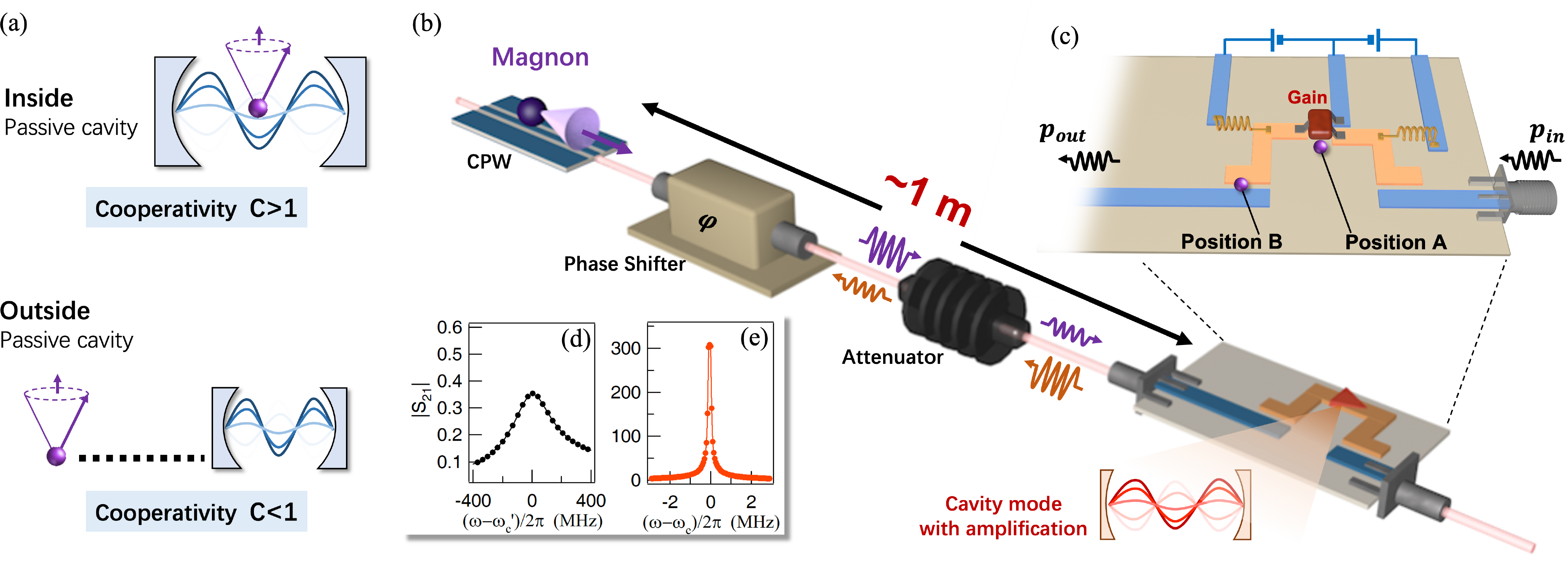,width=17.5cm}
\caption{ (a) Top: Strong photon-magnon coupling achieved in a passive cavity because of the direct mode overlap. Bottom: Photon-magnon cooperativity smaller than unity \cite{SM}, when the distance between subsystems is extended, hindering the realization of strong coupling. (b) Schematic picture of our experimental setup, where the cavity is a planar structure. A phase shifter and an attenuator are inserted between the cable and the cavity. They are used to modulate traveling waves in the coaxial cable. (c) Planar active cavity embedded with an amplifier. Positions A and B are selected to demonstrate strongly \textit{coherent} and \textit{dissipative} photon-magnon coupling, respectively. (d) and (e) Measured transmission spectra ($|S_{21}|$) of the cavity with the transistor off and on, respectively. When the transistor is on, the cavity mode shifts from $\omega_c'/2\pi=4.0$ GHz to $\omega_c/2\pi=3.82$ GHz.}\label{Fig1}
\end{center}
\end{figure*}

Among all possible methods, the most straightforward and simplest one is to use traveling waves in a waveguide to transfer interactions between separate systems \cite{van2013photon,hsu2016bound}. Such indirect coupling attracts attention because of its advantages in engineering abnormal dispersion and harnessing loss. Although enhancing subsystems' radiation to traveling waves can improve the indirect coupling strength between them, their dissipation will also increase. This constraint hinders the cooperativity of subsystems from surpassing unity \cite{SM} as sketched in Fig. \ref{Fig1} (a), thereby impeding the possibility of achieving strong coupling via this approach. To lift this limitation, we must solve the dilemma between increasing radiation and suppressing dissipation, thus enabling the strong coupling effect beyond the near field, even to meter-scale. 

Following this strategy, we design a cavity magnonic system \cite{rameshti2022cavity,zhang2014strongly,huebl2013high,goryachev2014high,zhang2015magnon,wang2019nonreciprocity,wang2021magnonic} to demonstrate the meter-scale strong coupling between microwave photons in an active cavity and magnons (i.e., energy quanta of spin waves) in a ferrimagnetic crystal. Recent work has demonstrated the generation of gain-driven polaritons with remarkable coherence in such a system \cite{PhysRevLett.130.146702}. However, the topic of long-range photon-magnon coupling that is tailored for an active cavity magnonic system has not yet been studied. The resonance of an active cavity is governed by the Van der Pol mechanism \cite{van1926,dutta2019critical,lorch2016genuine}, which has become a textbook knowledge and widely exists in different self-sustained systems, such as heartbeats \cite{van1928lxxii}, semiconductor lasers \cite{uemukai2000tunable} and trapped ions 
\cite{leibfried2003quantum}. Its core mechanism is a saturable gain, which can adaptively compensate for the cavity dissipation but doesn't suppress the cavity's radiation to traveling waves. By means of this merit, we are able to achieve the long-range strong coupling between the cavity photon mode and the magnon mode through manipulating their cooperative radiation, which was previously a challenge. The photon-magnon cooperativity is significantly enhanced, enabling both the strongly coherent and dissipative photon-magnon coupling at the meter-scale. Our work demonstrates a handy method for manipulating the strong coupling between distinct systems beyond the near-field regime, which may allow for constructing networks based on cavity magnonics for information processing tasks.

Figure \ref{Fig1} (b) depicts our experiment setup. The long-range strong coupling between microwave photons in an active cavity and magnons in a yttrium iron garnet (YIG) sphere is sustained by traveling waves in a long coaxial cable. The cavity is fabricated on a printed circuit board with FR-4 substrate. Its dimensions are $66\times40\times1.5$ mm$^3$. The FR-4 substrate has a dielectric constant of 4.2 and a loss tangent of 0.02. A low-noise transistor is embedded in the cavity (Fig. \ref{Fig1} (c)). All experiments were performed at room temperature with the amplifier on. The cavity photon mode can be modelled by a Hamiltonian $\mathcal{H}_a$ \cite{SM,gilles1993two,lee2013quantum}, which includes a saturable gain $-N\left[1-i\sqrt{\varepsilon/2}\hat{a}-\varepsilon|\hat{a}|^2\right]$, where $\hat{a}$, $N$ and $\varepsilon$ represent the photon operator, linear negative damping and saturation factor of the cavity mode, respectively. Because of the energy compensation from this saturable gain, the cavity mode is self-sustained. From the transmission spectra (Figs. \ref{Fig1} (d) and (e)), we see that the broad cavity resonance becomes an ultrasharp peak with an effective quality factor of $2.8\times10^4$ after turning on the transistor. To avoid perturbing the sensitive cavity mode, the probe signal is set to -50 dBm. The cavity mode frequency and its intrinsic damping extracted from curve fitting are $\omega_c/2\pi=3.82$~GHz and $\beta/2\pi=85.4$ MHz, respectively. The YIG sphere is mounted on a coplanar waveguide (CPW), and then connected to the cavity via a coaxial cable ($\sim1$ m) to produce the long-range strong photon-magnon coupling. The CPW is fabricated on a $22\times 50$ mm$^2$ RO4350B board. The width of its central strip is 1.1~mm and the gap between the central strip and the ground planes is 0.24 mm. The magnon mode in the sphere is modeled by a Hamiltonian $\mathcal{H}_m=\hbar (\omega_m-i\alpha)\hat{m}^{\dagger} \hat{m}$, where $\hat{m}$ ($\hat{m}^{\dagger}$) is the annihilation (creation) magnon operator, $\alpha$ is the magnon damping rate, and $\omega_m$ is the magnon mode frequency, which is precisely controlled by an external magnetic field. 

\begin{figure} [!htbp]
\begin{center}
\epsfig{file=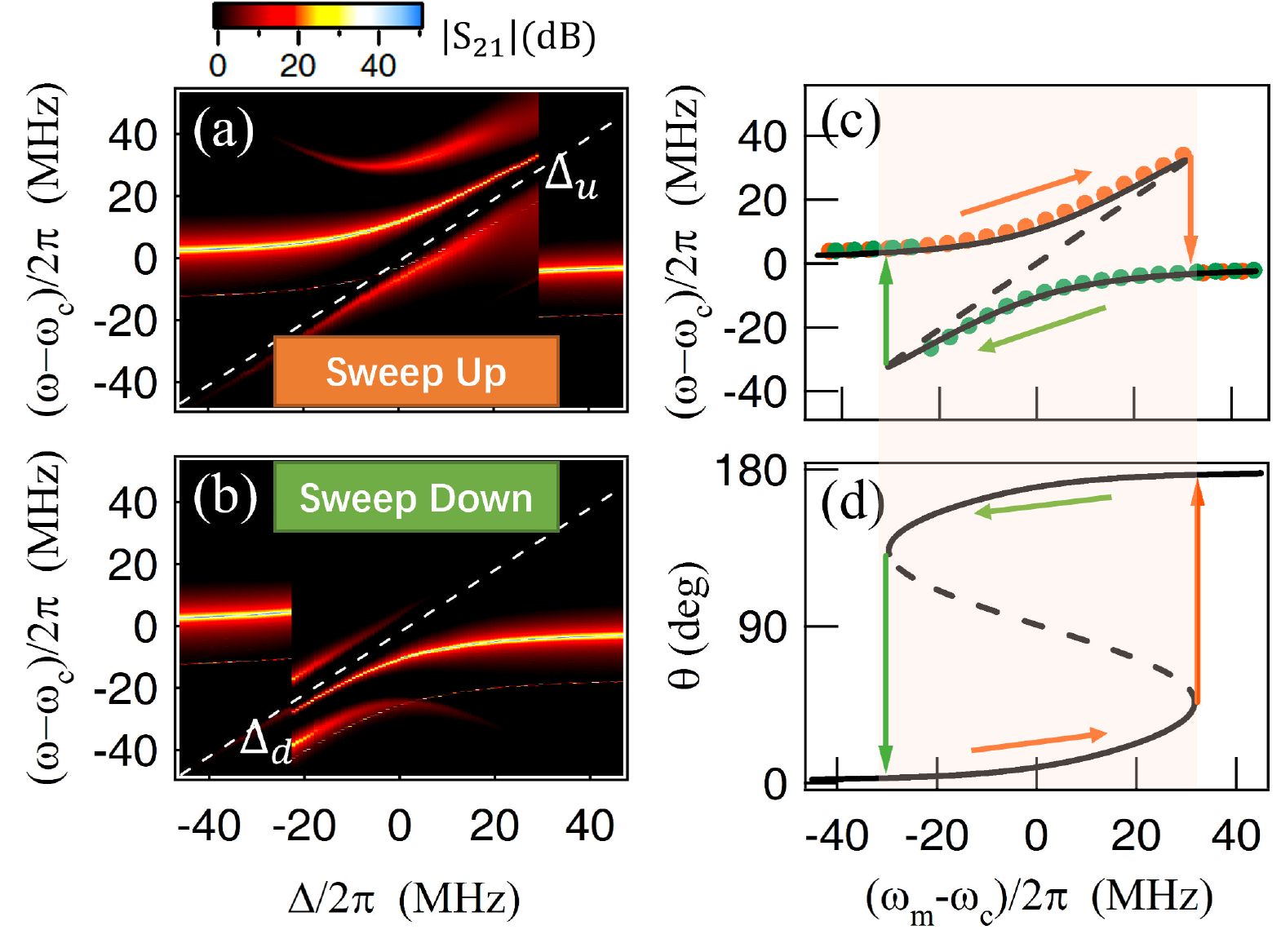,width=8.5cm}
\caption{ (a) and (b) Measured transmission spectra of our system with coherent photon-magnon coupling, when sweeping up and down the field detuning ($\Delta$), respectively. Two frequency jump points respectively occur at $\Delta_u$ and $\Delta_d$. White dashed lines indicate the dispersion of the magnon mode. (c) Frequencies of the synchronization mode extracted from (a) and (b) are plotted as a functions of $\Delta$. (d) Calculated relative phase ($\theta$) between two modes at different $\Delta$ values. The black lines are theoretical calculations by using Eq. (\ref{R_P}) and (\ref{VdP_Mode}). The dashed lines in (c) and (d) represent the unstable values of $\theta$ and $\omega_s$. Orange and green arrows indicate the processes of sweeping up and down $\Delta$.}\label{Fig2}
\end{center}
\end{figure}

Traveling waves propagating in the coaxial cable mediate the indirect photon-magnon coupling \cite{rao2020interactions}. The energy radiated by the magnon mode into traveling waves eventually converts to a perturbation on the cavity mode with a strength of $\sqrt{\kappa\gamma}e^{i\varphi}$, where $\varphi$ is the propagation phase of traveling waves between two separated modes, $\kappa/2\pi=18.7$ MHz (fitted from Fig. \ref{Fig1} (d)) and $\gamma$ (varied in the \textit{rf}-field \cite{yang2018influence}) are the radiation rates of two modes into traveling waves, respectively. Conversely, the perturbation on the magnon mode exerted by the cavity mode has a strength of $-\sqrt{\kappa\gamma}e^{i\varphi}$. This indirect coupling effect is a combination of coherent coupling with a strength $J=-\sqrt{\kappa\gamma}\sin\varphi$ and dissipative coupling with a strength $i\Gamma=i\sqrt{\kappa\gamma}\cos\varphi$. By tuning $\varphi$, the proportions of these two coupling mechanisms can be precisely controlled. 

The saturable gain compensates for the damping rate of the cavity mode to zero, but doesn't reduce the indirect photon-magnon coupling strength. This is the prerequisite for achieving the long-range strong photon-magnon coupling. Moreover, the coupling strength must exceed the overall damping rate of the magnon mode $\alpha'=\alpha+\gamma e^{i2\varphi}$. Therefore, the criteria for strong photon-magnon coupling become $\sqrt{\kappa\gamma}>\alpha'$, particularly $|J|>\alpha'$ and $|\Gamma|>\alpha'$ respectively for the coherent and dissipative coupling cases. By focusing on this goal, our experiment is designed in two steps: (i) preparing the strong coupling in the near-field regime, (ii) further pushing the distance limit of strong coupling to meter scale by manipulating traveling waves. 

First, we place the YIG sphere at the position A (Fig. \ref{Fig1} (c)), where the \textit{rf}-magnetic field of the cavity reaches its maximum. The direct overlap of the cavity and magnon modes produces a strong coherent coupling with a strength $g$ ($g\in\mathbb{R}$). The coupling between the magnon mode and the traveling waves is negligible at this position, thereby this purely coherent coupling is modeled as $\mathcal{H}_c=\hbar g(\hat{a}^{\dagger}+\hat{a})(\hat{m}^{\dagger}+\hat{m})$. By sweeping the field detuning ($\Delta=\omega_m-\omega_c$) up and down through tuning external magnetic field, the transmission spectra of the system are collected, as shown in Fig. \ref{Fig2} (a) and (b). In addition to weak side-band modes \cite{li2019nutation, SM}, the system possesses a dominant peak (bright yellow color). With different sweeping histories, this peak exhibits a bistable behavior. Two frequency jumps occur at $\Delta_u/2\pi=29.5$~MHz in Fig. \ref{Fig2}(a) and $\Delta_d/2\pi=-23.6$~MHz in Fig. \ref{Fig2}(b), which are almost symmetric with respect to the zero detuning ($\Delta=0$).

The dominant peak is a synchronization mode of the photon-magnon coupling. It can be explained as follows: the self-sustained oscillation of the cavity mode drives the spins to precess with a constant phase delay, while at the same time, the spin precession alters the susceptibility in the cavity field and causes a shift in the cavity oscillation frequency. Such a synchronization process is governed by the dynamic equations derived from system's Hamiltonian. After reaching the steady state, both the photon and magnon modes oscillate at the frequency of the synchronization mode $\omega_s$ with a constant phase difference $\theta$, i.e., $d\theta/dt=0$. Their oscillations can be written as $\hat{a}=|\hat{a}|e^{-i\omega_st}$ and $\hat{m}=|\hat{m}|e^{-i(\omega_st+\theta)}$. From the coupled dynamic equations \cite{SM}, we obtain
\begin{subequations}
\begin{align}
&(g\cos\theta-J\sin\theta)[\frac{|\hat{m}|}{|\hat{a}|}-\frac{|\hat{a}|}{|\hat{m}|}]+\Gamma\cos\theta[\frac{|\hat{m}|}{|\hat{a}|}+\frac{|\hat{a}|}{|\hat{m}|}]=\Delta \label{R_P}, \\
&\omega_s=\omega_c+[(g+\Gamma)\cos\theta-J\sin\theta]\frac{|\hat{m}|}{|\hat{a}|} \label{VdP_Mode},
\end{align}
\end{subequations}
where $|\hat{m}|/|\hat{a}|=\left[(g-\Gamma)\sin\theta+J\cos\theta\right]/\alpha'$ and $\theta\in\left(\arctan(-\frac{J}{g-\Gamma}),\quad\arctan(-\frac{J}{g-\Gamma})+\pi\right)$. The synchronization mode frequency $\omega_s$ is solely determined by the detuning ($\Delta$), coupling strength ($g$, $J+i\Gamma$) and magnon damping ($\alpha'$).

The synchronization mode frequencies extracted from Figs. \ref{Fig2} (a) and (b) are plotted in Fig. \ref{Fig2} (c). A clockwise hysteresis loop is formed between $\Delta_d$ and $\Delta_u$. Because the magnon mode decouples with traveling waves in transmission line, $J$ and $\Gamma$ are zero. $g/2\pi$ and $\alpha'/2\pi$ as $11$ MHz and $1.8$ MHz are fitted from the measured $\omega_s$ in Fig. \ref{Fig2} (c) by using Eq. (\ref{VdP_Mode}). Because $g>\alpha'$, strongly coherent photon-magnon coupling is realized. The calculated $\theta$ is an S-shape function of $\Delta$ (Fig. \ref{Fig2} (d)). Between $\Delta_d$ and $\Delta_u$, $\theta$ has three analytical solutions, including an unstable solution indicated by a dashed line. At $\Delta_d$ and $\Delta_u$, $\theta$ reaches its value boundaries and exhibits abrupt jumps\cite{pippard2007physics}.

\begin{figure} [!htbp]
\begin{center}
\epsfig{file=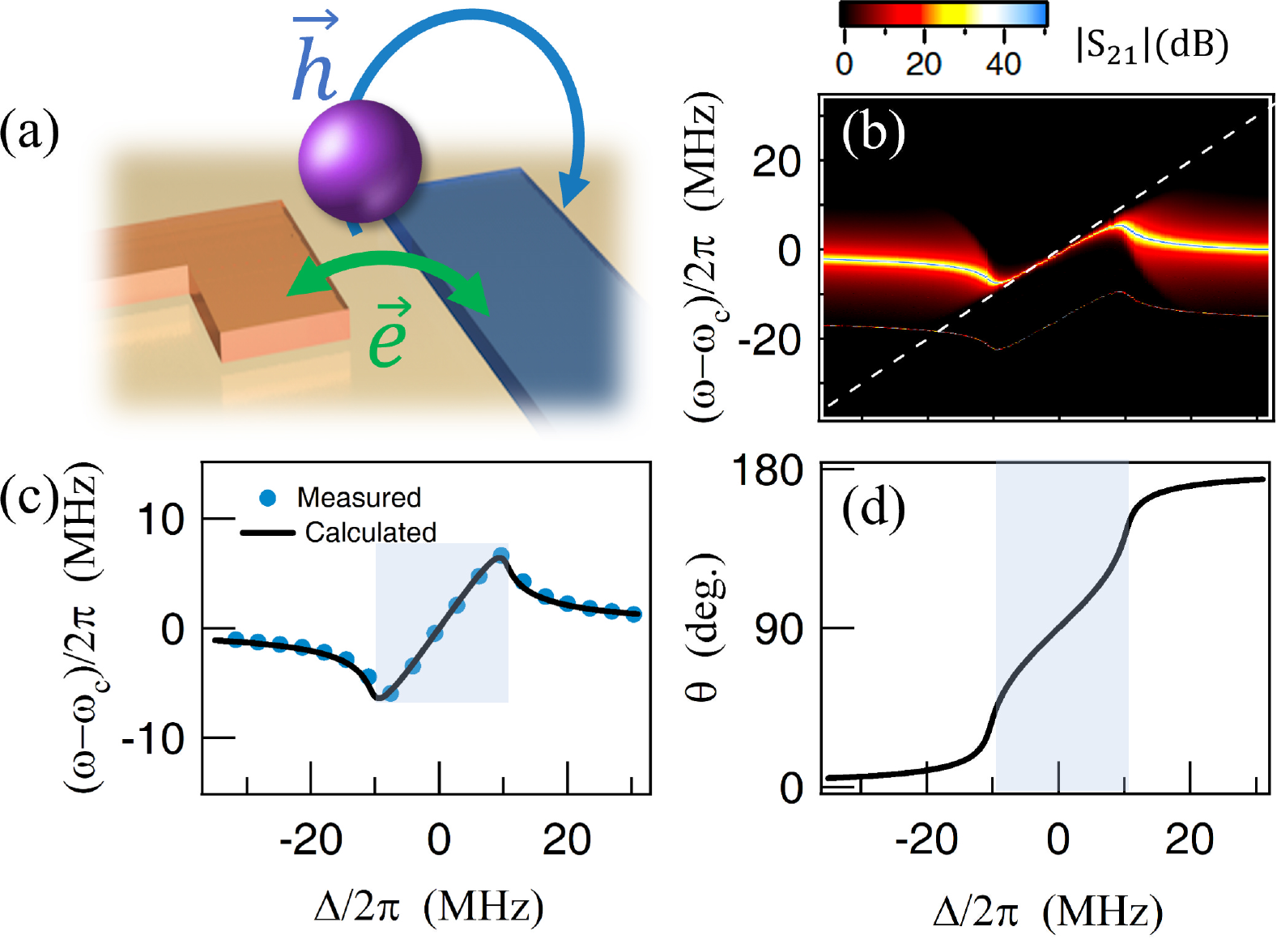,width=8.5cm}
\caption{ (a) YIG sphere position for realizing the strongly dissipative photon-magnon coupling. $\vec{h}$ and $\vec{e}$ represent the microwave magnetic and electric fields, respectively. (b) Measured transmission spectra of our system with dissipative photon-magnon coupling at different detuning. (c) Extracted synchronization mode frequencies from (b) plotted as functions of $\Delta$. The solid line is the theoretical calculation. (d) Calculated $\theta$ at different $\Delta$ values by using Eq. (\ref{R_P}). Blue color in (c) and (d) indicates the detuning range of the mode coalescence.}\label{DissCoup}
\end{center}
\end{figure}

To achieve strongly dissipative photon-magnon coupling, we must use the mediation of traveling waves. For this purpose, we place the YIG sphere near the output stripline of the cavity (Fig. \ref{DissCoup} (a)). In this configuration, the direct coherent coupling $g$ is suppressed, while the indirect dissipative coupling is sustained by traveling waves in the output stripline. Unlike the coherent coupling case, the bistability of the synchronization mode disappears, as shown in Fig. \ref{DissCoup} (b); nevertheless, a level attraction with the coalescence of two hybridized photon-magnon modes occurs instead \cite{rao2021interferometric,grigoryan2018synchronized,yu2019prediction,bhoi2019abnormal,boventer2019steering,harder2018level}. It demonstrates the realization of the dissipative photon-magnon coupling. This dissipative coupling is inaccessible to the strong coupling regime in passive cavity magnonic systems, because the cooperative radiation \cite{wang2019nonreciprocity, rao2020interactions} that produces the dissipative coupling limits the photon-magnon cooperativity to less than unity. However, in our case, with fitting the dispersion trace (blue circles in Fig. \ref{DissCoup} (c)), $\Gamma/2\pi=6.2$ MHz, $\alpha'/2\pi=3$ MHz are obtained. $g/2\pi$ is approximately zero. Although the magnon mode damping is enhanced by the radiation into traveling waves (verified by far-detuned curve fitting \cite{SM}), the dissipative coupling strength still exceeds magnon damping ($\Gamma>\alpha'$), unambiguously demonstrating the realization of the strongly dissipative coupling. In addition, $\theta$ calculated from Eq. (\ref{R_P}) reduces from an S-shape function to a monotonic function (see Fig. \ref{DissCoup} (d)), causing the bistability to disappear. 

\begin{figure} [!htbp]
\begin{center}
\epsfig{file=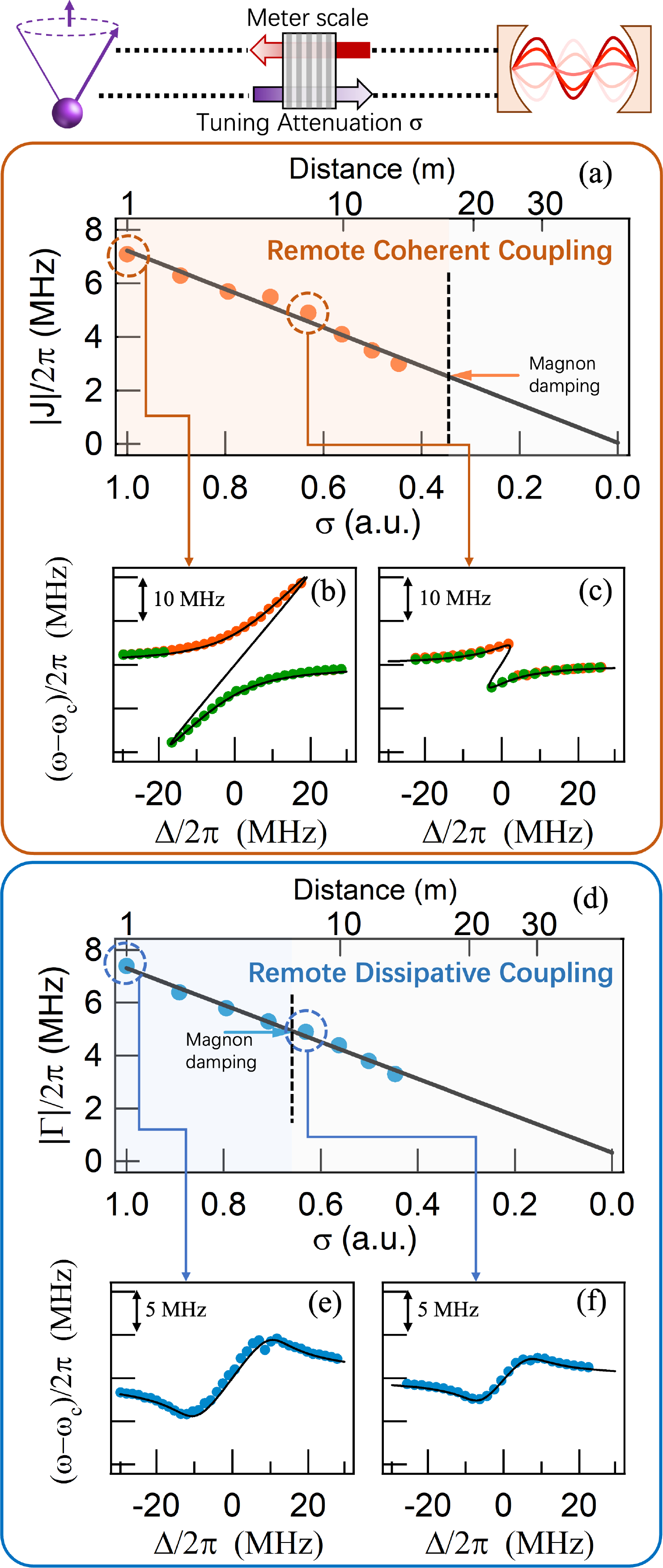,width=8cm}
\caption{ (a) and (d) Coherent and dissipative coupling strengths as linear functions of the transmission coefficient $\sigma$ of the attenuator. Their top axes indicate the equivalent length of coaxial cable for each $\sigma$. Black solid lines are linear fitting results. The orange (blue) arrow indicates the intersection of the coherent (dissipative) coupling strength and the magnon damping. (b) and (c) Bistability of $\omega_s$ measured with $\sigma=1$ and $0.63$. Orange and green dots represent synchronization mode frequencies from upward and downward sweep of $\Delta$. (e) and (f) Measured $\omega_s$ in the purely dissipative coupling case with $\sigma=1$ and $0.63$. Black solid lines in (b), (c) , (e), (f) are calculated by using Eq. (\ref{VdP_Mode}).}\label{remote}
\end{center}
\end{figure}

By moving the YIG sphere further along the output stripline of the cavity, the propagation phase of traveling waves $\varphi$ can be precisely tuned. As a result, the indirect photon-magnon coupling becomes a combination of coherent and dissipative couplings with a strength $J+i\Gamma$. This effect has been verified in experiment, and can be well explained by Eq. (\ref{VdP_Mode}) \cite{SM}. It demonstrates the feasibility of achieving long-range photon-magnon coupling by utilizing traveling waves as a mediator.

Following this idea, we then extend the spatial distance between the cavity and the YIG sphere to the meter scale (Fig. \ref{Fig1} (a)) via a coaxial cable ($\sim 1$ m). With full access to traveling waves, we can control the strong photon-magnon coupling without manipulating two subsystems. The voltage-controlled attenuator and the mechanical phase shifter modulate the transmission coefficient ($\sigma$) and the propagation phase ($\varphi$) of traveling waves, respectively. First, we form purely coherent photon-magnon coupling by adjusting $\varphi$. When the bias voltage on the attenuator is $0.2$~V, its transmission $\sigma$ is almost 0 dB. The coherent coupling strength and the magnon damping fitted from measured $\omega_s$ are $|J|/2\pi=7.1$ MHz and $\alpha'/2\pi=1.3$ MHz, respectively, satisfying the criterion $|J|>\alpha'$ for strongly coherent coupling. As the attenuator is tuned to further decrease $\sigma$, the coherent coupling strength ($|J|$) linearly decays (Fig. \ref{remote} (a)). The strong coupling status can be maintained even at $\sigma=0.35$. Considering that the attenuation of the coaxial cable is 0.56 dB/m, the longest distance for strongly coherent photon-magnon coupling is estimated as 17.3~m. To support this statement, a real-cable experiment was conducted. This cable has a lower attenuation rate than the 1-meter cable, and hence allows us to achieve a strong coherent coupling of 20 m \cite{SM}. The strength almost decreases to zero at $\sigma=0$, corresponding to photon-magnon decoupling when the cable transmission is shut off. Two measurement results with $\sigma=1$ and $0.63$ are plotted in Fig. \ref{remote} (b) and (c), from which both the coupling gap and the bistability range shrink with decreasing $|J|$.

Then, we add $\pi/2$ phase to travelling waves by adjusting the phase shifter. Consequently, photon-magnon coupling is switched to be purely dissipative. With increasing attenuation, the dissipative coupling strength ($|\Gamma|$) linearly decreases (Fig. \ref{remote} (d)), similar to the aforementioned coherent coupling case. When $\sigma=1$, the dissipative coupling strength ($|\Gamma|/2\pi$) and the magnon damping ($\alpha'/2\pi$) are fitted as 7.4 MHz and 6.2 MHz, respectively. It still meets the criterion of $|\Gamma|>\alpha'$. Hence, strongly dissipative photon-magnon coupling over meter scale is demonstrated. Two typical measurements are plotted in Fig. \ref{remote} (e) and (f). We note that the increase of magnon damping in the dissipative coupling hinders the cooperativity, and the equivalent strong coupling distance estimated is 7.6~m. From Fig. \ref{remote} (e) to (f), the detuning range of the mode coalescence decreases.

Our results demonstrate two distinct kinds of strong photon-magnon coupling effects, namely the coherent and dissipative coupling, which can be extended beyond the near-field limitation. Different from previous studies that control photon-magnon coupling strength by changing spin number, cavity mode volume, or mode overlap coefficient, we demonstrate a new method for controlling photon-magnon coupling by manipulating their common radiative mediums, i.e., traveling waves. By tuning the propagation phase of traveling waves, the switch between strongly coherent and dissipative photon-magnon coupling is realized. By controlling the transmission coefficient of traveling waves,
the photon-magnon coupling can be precisely controlled without manipulating the subsystems themselves. It is reasonable to believe that the coupling distance can be further enhanced by amplifying traveling waves, thus providing further opportunities for large-scale quantum network and communication experiments. It allows magnons to exchange and access information between spatially separated and physically distinct systems in a network architecture. Magnonic functionalities, such as long-lived memories \cite{zhang2015magnon}, nonreciprocal transmission \cite{wang2019nonreciprocity}, logic operations \cite{rao2019analogue} and energy accumulation \cite{yu2020magnon}, which originally demonstrated in a single device, are expected to be integrated in a large scale network. Our results demonstrate a generalized method for achieving long-range strong coupling and promise new approaches for information processing in cavity magnonic networks.

\acknowledgements

This work has been funded by National Natural Science Foundation of China under Grants Nos.12122413, 11974369, 11991063 and 12204306, STCSM Nos.21JC1406200 and 22JC1403300, the Youth Innovation Promotion Association  No. 2020247 and Strategic priority research No. XDB43010200 of CAS, the National Key R\&D Program of China (No. 2022YFA1404603, 2022YFA1604400), the SITP Independent Foundation, the Shanghai Pujiang Program (No. 22PJ1410700). We thank C.-M. Hu for his suggestion.


\begin{thebibliography}{49}%
\makeatletter
\providecommand \@ifxundefined [1]{%
 \@ifx{#1\undefined}
}%
\providecommand \@ifnum [1]{%
 \ifnum #1\expandafter \@firstoftwo
 \else \expandafter \@secondoftwo
 \fi
}%
\providecommand \@ifx [1]{%
 \ifx #1\expandafter \@firstoftwo
 \else \expandafter \@secondoftwo
 \fi
}%
\providecommand \natexlab [1]{#1}%
\providecommand \enquote  [1]{``#1''}%
\providecommand \bibnamefont  [1]{#1}%
\providecommand \bibfnamefont [1]{#1}%
\providecommand \citenamefont [1]{#1}%
\providecommand \href@noop [0]{\@secondoftwo}%
\providecommand \href [0]{\begingroup \@sanitize@url \@href}%
\providecommand \@href[1]{\@@startlink{#1}\@@href}%
\providecommand \@@href[1]{\endgroup#1\@@endlink}%
\providecommand \@sanitize@url [0]{\catcode `\\12\catcode `\$12\catcode
  `\&12\catcode `\#12\catcode `\^12\catcode `\_12\catcode `\%12\relax}%
\providecommand \@@startlink[1]{}%
\providecommand \@@endlink[0]{}%
\providecommand \url  [0]{\begingroup\@sanitize@url \@url }%
\providecommand \@url [1]{\endgroup\@href {#1}{\urlprefix }}%
\providecommand \urlprefix  [0]{URL }%
\providecommand \Eprint [0]{\href }%
\providecommand \doibase [0]{http://dx.doi.org/}%
\providecommand \selectlanguage [0]{\@gobble}%
\providecommand \bibinfo  [0]{\@secondoftwo}%
\providecommand \bibfield  [0]{\@secondoftwo}%
\providecommand \translation [1]{[#1]}%
\providecommand \BibitemOpen [0]{}%
\providecommand \bibitemStop [0]{}%
\providecommand \bibitemNoStop [0]{.\EOS\space}%
\providecommand \EOS [0]{\spacefactor3000\relax}%
\providecommand \BibitemShut  [1]{\csname bibitem#1\endcsname}%
\let\auto@bib@innerbib\@empty
%</preamble>
\bibitem [{\citenamefont {Pezze}\ \emph {et~al.}(2018)\citenamefont {Pezze},
  \citenamefont {Smerzi}, \citenamefont {Oberthaler}, \citenamefont {Schmied},\
  and\ \citenamefont {Treutlein}}]{pezze2018quantum}%
  \BibitemOpen
  \bibfield  {author} {\bibinfo {author} {\bibfnamefont {L.}~\bibnamefont
  {Pezze}}, \bibinfo {author} {\bibfnamefont {A.}~\bibnamefont {Smerzi}},
  \bibinfo {author} {\bibfnamefont {M.~K.}\ \bibnamefont {Oberthaler}},
  \bibinfo {author} {\bibfnamefont {R.}~\bibnamefont {Schmied}}, \ and\
  \bibinfo {author} {\bibfnamefont {P.}~\bibnamefont {Treutlein}},\ }\bibfield
  {title} {Quantum metrology with nonclassical states of atomic ensembles,\
  }\href@noop {} {\bibfield  {journal} {\bibinfo  {journal} {Reviews of Modern
  Physics}\ }\textbf {\bibinfo {volume} {90}},\ \bibinfo {pages} {035005}
  (\bibinfo {year} {2018})}\BibitemShut {NoStop}%
\bibitem [{\citenamefont {Gross}\ and\ \citenamefont
  {Bloch}(2017)}]{gross2017quantum}%
  \BibitemOpen
  \bibfield  {author} {\bibinfo {author} {\bibfnamefont {C.}~\bibnamefont
  {Gross}}\ and\ \bibinfo {author} {\bibfnamefont {I.}~\bibnamefont {Bloch}},\
  }\bibfield  {title} {Quantum simulations with ultracold atoms in optical
  lattices,\ }\href@noop {} {\bibfield  {journal} {\bibinfo  {journal}
  {Science}\ }\textbf {\bibinfo {volume} {357}},\ \bibinfo {pages} {995}
  (\bibinfo {year} {2017})}\BibitemShut {NoStop}%
\bibitem [{\citenamefont {Kimble}(2008)}]{kimble2008quantum}%
  \BibitemOpen
  \bibfield  {author} {\bibinfo {author} {\bibfnamefont {H.~J.}\ \bibnamefont
  {Kimble}},\ }\bibfield  {title} {The quantum internet,\ }\href@noop {}
  {\bibfield  {journal} {\bibinfo  {journal} {Nature}\ }\textbf {\bibinfo
  {volume} {453}},\ \bibinfo {pages} {1023} (\bibinfo {year}
  {2008})}\BibitemShut {NoStop}%
\bibitem [{\citenamefont {Tabuchi}\ \emph {et~al.}(2015)\citenamefont
  {Tabuchi}, \citenamefont {Ishino}, \citenamefont {Noguchi}, \citenamefont
  {Ishikawa}, \citenamefont {Yamazaki}, \citenamefont {Usami},\ and\
  \citenamefont {Nakamura}}]{tabuchi2015coherent}%
  \BibitemOpen
  \bibfield  {author} {\bibinfo {author} {\bibfnamefont {Y.}~\bibnamefont
  {Tabuchi}}, \bibinfo {author} {\bibfnamefont {S.}~\bibnamefont {Ishino}},
  \bibinfo {author} {\bibfnamefont {A.}~\bibnamefont {Noguchi}}, \bibinfo
  {author} {\bibfnamefont {T.}~\bibnamefont {Ishikawa}}, \bibinfo {author}
  {\bibfnamefont {R.}~\bibnamefont {Yamazaki}}, \bibinfo {author}
  {\bibfnamefont {K.}~\bibnamefont {Usami}}, \ and\ \bibinfo {author}
  {\bibfnamefont {Y.}~\bibnamefont {Nakamura}},\ }\bibfield  {title} {Coherent
  coupling between a ferromagnetic magnon and a superconducting qubit,\
  }\href@noop {} {\bibfield  {journal} {\bibinfo  {journal} {Science}\ }\textbf
  {\bibinfo {volume} {349}},\ \bibinfo {pages} {405} (\bibinfo {year}
  {2015})}\BibitemShut {NoStop}%
\bibitem [{\citenamefont {Lachance-Quirion}\ \emph {et~al.}(2020)\citenamefont
  {Lachance-Quirion}, \citenamefont {Wolski}, \citenamefont {Tabuchi},
  \citenamefont {Kono}, \citenamefont {Usami},\ and\ \citenamefont
  {Nakamura}}]{lachance2020entanglement}%
  \BibitemOpen
  \bibfield  {author} {\bibinfo {author} {\bibfnamefont {D.}~\bibnamefont
  {Lachance-Quirion}}, \bibinfo {author} {\bibfnamefont {S.~P.}\ \bibnamefont
  {Wolski}}, \bibinfo {author} {\bibfnamefont {Y.}~\bibnamefont {Tabuchi}},
  \bibinfo {author} {\bibfnamefont {S.}~\bibnamefont {Kono}}, \bibinfo {author}
  {\bibfnamefont {K.}~\bibnamefont {Usami}}, \ and\ \bibinfo {author}
  {\bibfnamefont {Y.}~\bibnamefont {Nakamura}},\ }\bibfield  {title}
  {Entanglement-based single-shot detection of a single magnon with a
  superconducting qubit,\ }\href@noop {} {\bibfield  {journal} {\bibinfo
  {journal} {Science}\ }\textbf {\bibinfo {volume} {367}},\ \bibinfo {pages}
  {425} (\bibinfo {year} {2020})}\BibitemShut {NoStop}%
\bibitem [{\citenamefont {Gibbs}\ \emph {et~al.}(2011)\citenamefont {Gibbs},
  \citenamefont {Khitrova},\ and\ \citenamefont {Koch}}]{gibbs2011exciton}%
  \BibitemOpen
  \bibfield  {author} {\bibinfo {author} {\bibfnamefont {H.}~\bibnamefont
  {Gibbs}}, \bibinfo {author} {\bibfnamefont {G.}~\bibnamefont {Khitrova}}, \
  and\ \bibinfo {author} {\bibfnamefont {S.}~\bibnamefont {Koch}},\ }\bibfield
  {title} {Exciton--polariton light--semiconductor coupling effects,\
  }\href@noop {} {\bibfield  {journal} {\bibinfo  {journal} {Nature Photonics}\
  }\textbf {\bibinfo {volume} {5}},\ \bibinfo {pages} {273} (\bibinfo {year}
  {2011})}\BibitemShut {NoStop}%
\bibitem [{\citenamefont {Mabuchi}\ and\ \citenamefont
  {Doherty}(2002)}]{mabuchi2002cavity}%
  \BibitemOpen
  \bibfield  {author} {\bibinfo {author} {\bibfnamefont {H.}~\bibnamefont
  {Mabuchi}}\ and\ \bibinfo {author} {\bibfnamefont {A.}~\bibnamefont
  {Doherty}},\ }\bibfield  {title} {Cavity quantum electrodynamics: coherence
  in context,\ }\href@noop {} {\bibfield  {journal} {\bibinfo  {journal}
  {Science}\ }\textbf {\bibinfo {volume} {298}},\ \bibinfo {pages} {1372}
  (\bibinfo {year} {2002})}\BibitemShut {NoStop}%
\bibitem [{\citenamefont {Scully}\ and\ \citenamefont
  {Zubairy}(1999)}]{scully1999quantum}%
  \BibitemOpen
  \bibfield  {author} {\bibinfo {author} {\bibfnamefont {M.~O.}\ \bibnamefont
  {Scully}}\ and\ \bibinfo {author} {\bibfnamefont {M.~S.}\ \bibnamefont
  {Zubairy}},\ }\href@noop {} {Quantum optics} (\bibinfo {year}
  {1999})\BibitemShut {NoStop}%
\bibitem [{\citenamefont {Richerme}\ \emph {et~al.}(2014)\citenamefont
  {Richerme}, \citenamefont {Gong}, \citenamefont {Lee}, \citenamefont {Senko},
  \citenamefont {Smith}, \citenamefont {Foss-Feig}, \citenamefont {Michalakis},
  \citenamefont {Gorshkov},\ and\ \citenamefont {Monroe}}]{richerme2014non}%
  \BibitemOpen
  \bibfield  {author} {\bibinfo {author} {\bibfnamefont {P.}~\bibnamefont
  {Richerme}}, \bibinfo {author} {\bibfnamefont {Z.-X.}\ \bibnamefont {Gong}},
  \bibinfo {author} {\bibfnamefont {A.}~\bibnamefont {Lee}}, \bibinfo {author}
  {\bibfnamefont {C.}~\bibnamefont {Senko}}, \bibinfo {author} {\bibfnamefont
  {J.}~\bibnamefont {Smith}}, \bibinfo {author} {\bibfnamefont
  {M.}~\bibnamefont {Foss-Feig}}, \bibinfo {author} {\bibfnamefont
  {S.}~\bibnamefont {Michalakis}}, \bibinfo {author} {\bibfnamefont {A.~V.}\
  \bibnamefont {Gorshkov}}, \ and\ \bibinfo {author} {\bibfnamefont
  {C.}~\bibnamefont {Monroe}},\ }\bibfield  {title} {Non-local propagation of
  correlations in quantum systems with long-range interactions,\ }\href@noop {}
  {\bibfield  {journal} {\bibinfo  {journal} {Nature}\ }\textbf {\bibinfo
  {volume} {511}},\ \bibinfo {pages} {198} (\bibinfo {year}
  {2014})}\BibitemShut {NoStop}%
\bibitem [{\citenamefont {Landig}\ \emph {et~al.}(2016)\citenamefont {Landig},
  \citenamefont {Hruby}, \citenamefont {Dogra}, \citenamefont {Landini},
  \citenamefont {Mottl}, \citenamefont {Donner},\ and\ \citenamefont
  {Esslinger}}]{landig2016quantum}%
  \BibitemOpen
  \bibfield  {author} {\bibinfo {author} {\bibfnamefont {R.}~\bibnamefont
  {Landig}}, \bibinfo {author} {\bibfnamefont {L.}~\bibnamefont {Hruby}},
  \bibinfo {author} {\bibfnamefont {N.}~\bibnamefont {Dogra}}, \bibinfo
  {author} {\bibfnamefont {M.}~\bibnamefont {Landini}}, \bibinfo {author}
  {\bibfnamefont {R.}~\bibnamefont {Mottl}}, \bibinfo {author} {\bibfnamefont
  {T.}~\bibnamefont {Donner}}, \ and\ \bibinfo {author} {\bibfnamefont
  {T.}~\bibnamefont {Esslinger}},\ }\bibfield  {title} {Quantum phases from
  competing short-and long-range interactions in an optical lattice,\
  }\href@noop {} {\bibfield  {journal} {\bibinfo  {journal} {Nature}\ }\textbf
  {\bibinfo {volume} {532}},\ \bibinfo {pages} {476} (\bibinfo {year}
  {2016})}\BibitemShut {NoStop}%
\bibitem [{\citenamefont {Mottl}\ \emph {et~al.}(2012)\citenamefont {Mottl},
  \citenamefont {Brennecke}, \citenamefont {Baumann}, \citenamefont {Landig},
  \citenamefont {Donner},\ and\ \citenamefont {Esslinger}}]{mottl2012roton}%
  \BibitemOpen
  \bibfield  {author} {\bibinfo {author} {\bibfnamefont {R.}~\bibnamefont
  {Mottl}}, \bibinfo {author} {\bibfnamefont {F.}~\bibnamefont {Brennecke}},
  \bibinfo {author} {\bibfnamefont {K.}~\bibnamefont {Baumann}}, \bibinfo
  {author} {\bibfnamefont {R.}~\bibnamefont {Landig}}, \bibinfo {author}
  {\bibfnamefont {T.}~\bibnamefont {Donner}}, \ and\ \bibinfo {author}
  {\bibfnamefont {T.}~\bibnamefont {Esslinger}},\ }\bibfield  {title}
  {Roton-type mode softening in a quantum gas with cavity-mediated long-range
  interactions,\ }\href@noop {} {\bibfield  {journal} {\bibinfo  {journal}
  {Science}\ }\textbf {\bibinfo {volume} {336}},\ \bibinfo {pages} {1570}
  (\bibinfo {year} {2012})}\BibitemShut {NoStop}%
\bibitem [{\citenamefont {Fukami}\ \emph {et~al.}(2021)\citenamefont {Fukami},
  \citenamefont {Candido}, \citenamefont {Awschalom},\ and\ \citenamefont
  {Flatt\'e}}]{PRXQuantum2040314}%
  \BibitemOpen
  \bibfield  {author} {\bibinfo {author} {\bibfnamefont {M.}~\bibnamefont
  {Fukami}}, \bibinfo {author} {\bibfnamefont {D.~R.}\ \bibnamefont {Candido}},
  \bibinfo {author} {\bibfnamefont {D.~D.}\ \bibnamefont {Awschalom}}, \ and\
  \bibinfo {author} {\bibfnamefont {M.~E.}\ \bibnamefont {Flatt\'e}},\
  }\bibfield  {title} {Opportunities for long-range magnon-mediated
  entanglement of spin qubits via on- and off-resonant coupling,\ }\href
  {\doibase 10.1103/PRXQuantum.2.040314} {\bibfield  {journal} {\bibinfo
  {journal} {PRX Quantum}\ }\textbf {\bibinfo {volume} {2}},\ \bibinfo {pages}
  {040314} (\bibinfo {year} {2021})}\BibitemShut {NoStop}%
\bibitem [{\citenamefont {Wang}\ \emph {et~al.}(2022)\citenamefont {Wang},
  \citenamefont {Wang}, \citenamefont {Yao}, \citenamefont {Shen},
  \citenamefont {Wu}, \citenamefont {Qian}, \citenamefont {Li}, \citenamefont
  {Zhu},\ and\ \citenamefont {You}}]{wang2022giant}%
  \BibitemOpen
  \bibfield  {author} {\bibinfo {author} {\bibfnamefont {Z.-Q.}\ \bibnamefont
  {Wang}}, \bibinfo {author} {\bibfnamefont {Y.-P.}\ \bibnamefont {Wang}},
  \bibinfo {author} {\bibfnamefont {J.}~\bibnamefont {Yao}}, \bibinfo {author}
  {\bibfnamefont {R.-C.}\ \bibnamefont {Shen}}, \bibinfo {author}
  {\bibfnamefont {W.-J.}\ \bibnamefont {Wu}}, \bibinfo {author} {\bibfnamefont
  {J.}~\bibnamefont {Qian}}, \bibinfo {author} {\bibfnamefont {J.}~\bibnamefont
  {Li}}, \bibinfo {author} {\bibfnamefont {S.-Y.}\ \bibnamefont {Zhu}}, \ and\
  \bibinfo {author} {\bibfnamefont {J.}~\bibnamefont {You}},\ }\bibfield
  {title} {Giant spin ensembles in waveguide magnonics,\ }\href@noop {}
  {\bibfield  {journal} {\bibinfo  {journal} {Nature Communications}\ }\textbf
  {\bibinfo {volume} {13}},\ \bibinfo {pages} {7580} (\bibinfo {year}
  {2022})}\BibitemShut {NoStop}%
\bibitem [{\citenamefont {Li}\ \emph {et~al.}(2022)\citenamefont {Li},
  \citenamefont {Yefremenko}, \citenamefont {Lisovenko}, \citenamefont
  {Trevillian}, \citenamefont {Polakovic}, \citenamefont {Cecil}, \citenamefont
  {Barry}, \citenamefont {Pearson}, \citenamefont {Divan}, \citenamefont
  {Tyberkevych} \emph {et~al.}}]{li2022coherent}%
  \BibitemOpen
  \bibfield  {author} {\bibinfo {author} {\bibfnamefont {Y.}~\bibnamefont
  {Li}}, \bibinfo {author} {\bibfnamefont {V.~G.}\ \bibnamefont {Yefremenko}},
  \bibinfo {author} {\bibfnamefont {M.}~\bibnamefont {Lisovenko}}, \bibinfo
  {author} {\bibfnamefont {C.}~\bibnamefont {Trevillian}}, \bibinfo {author}
  {\bibfnamefont {T.}~\bibnamefont {Polakovic}}, \bibinfo {author}
  {\bibfnamefont {T.~W.}\ \bibnamefont {Cecil}}, \bibinfo {author}
  {\bibfnamefont {P.~S.}\ \bibnamefont {Barry}}, \bibinfo {author}
  {\bibfnamefont {J.}~\bibnamefont {Pearson}}, \bibinfo {author} {\bibfnamefont
  {R.}~\bibnamefont {Divan}}, \bibinfo {author} {\bibfnamefont
  {V.}~\bibnamefont {Tyberkevych}},  \emph {et~al.},\ }\bibfield  {title}
  {Coherent coupling of two remote magnonic resonators mediated by
  superconducting circuits,\ }\href@noop {} {\bibfield  {journal} {\bibinfo
  {journal} {Phys. Rev. Lett.}\ }\textbf {\bibinfo {volume} {128}},\ \bibinfo
  {pages} {047701} (\bibinfo {year} {2022})}\BibitemShut {NoStop}%
\bibitem [{\citenamefont {Gopalakrishnan}\ \emph {et~al.}(2009)\citenamefont
  {Gopalakrishnan}, \citenamefont {Lev},\ and\ \citenamefont
  {Goldbart}}]{gopalakrishnan2009emergent}%
  \BibitemOpen
  \bibfield  {author} {\bibinfo {author} {\bibfnamefont {S.}~\bibnamefont
  {Gopalakrishnan}}, \bibinfo {author} {\bibfnamefont {B.~L.}\ \bibnamefont
  {Lev}}, \ and\ \bibinfo {author} {\bibfnamefont {P.~M.}\ \bibnamefont
  {Goldbart}},\ }\bibfield  {title} {Emergent crystallinity and frustration
  with bose--einstein condensates in multimode cavities,\ }\href@noop {}
  {\bibfield  {journal} {\bibinfo  {journal} {Nature Physics}\ }\textbf
  {\bibinfo {volume} {5}},\ \bibinfo {pages} {845} (\bibinfo {year}
  {2009})}\BibitemShut {NoStop}%
\bibitem [{\citenamefont {Maghrebi}\ \emph {et~al.}(2017)\citenamefont
  {Maghrebi}, \citenamefont {Gong},\ and\ \citenamefont
  {Gorshkov}}]{maghrebi2017continuous}%
  \BibitemOpen
  \bibfield  {author} {\bibinfo {author} {\bibfnamefont {M.~F.}\ \bibnamefont
  {Maghrebi}}, \bibinfo {author} {\bibfnamefont {Z.-X.}\ \bibnamefont {Gong}},
  \ and\ \bibinfo {author} {\bibfnamefont {A.~V.}\ \bibnamefont {Gorshkov}},\
  }\bibfield  {title} {Continuous symmetry breaking in 1d long-range
  interacting quantum systems,\ }\href@noop {} {\bibfield  {journal} {\bibinfo
  {journal} {Physical Review Letters}\ }\textbf {\bibinfo {volume} {119}},\
  \bibinfo {pages} {023001} (\bibinfo {year} {2017})}\BibitemShut {NoStop}%
\bibitem [{\citenamefont {Hofmann}\ \emph {et~al.}(2012)\citenamefont
  {Hofmann}, \citenamefont {Krug}, \citenamefont {Ortegel}, \citenamefont
  {G{\'e}rard}, \citenamefont {Weber}, \citenamefont {Rosenfeld},\ and\
  \citenamefont {Weinfurter}}]{hofmann2012heralded}%
  \BibitemOpen
  \bibfield  {author} {\bibinfo {author} {\bibfnamefont {J.}~\bibnamefont
  {Hofmann}}, \bibinfo {author} {\bibfnamefont {M.}~\bibnamefont {Krug}},
  \bibinfo {author} {\bibfnamefont {N.}~\bibnamefont {Ortegel}}, \bibinfo
  {author} {\bibfnamefont {L.}~\bibnamefont {G{\'e}rard}}, \bibinfo {author}
  {\bibfnamefont {M.}~\bibnamefont {Weber}}, \bibinfo {author} {\bibfnamefont
  {W.}~\bibnamefont {Rosenfeld}}, \ and\ \bibinfo {author} {\bibfnamefont
  {H.}~\bibnamefont {Weinfurter}},\ }\bibfield  {title} {Heralded entanglement
  between widely separated atoms,\ }\href@noop {} {\bibfield  {journal}
  {\bibinfo  {journal} {Science}\ }\textbf {\bibinfo {volume} {337}},\ \bibinfo
  {pages} {72} (\bibinfo {year} {2012})}\BibitemShut {NoStop}%
\bibitem [{\citenamefont {Karg}\ \emph {et~al.}(2020)\citenamefont {Karg},
  \citenamefont {Gouraud}, \citenamefont {Ngai}, \citenamefont {Schmid},
  \citenamefont {Hammerer},\ and\ \citenamefont {Treutlein}}]{karg2020light}%
  \BibitemOpen
  \bibfield  {author} {\bibinfo {author} {\bibfnamefont {T.~M.}\ \bibnamefont
  {Karg}}, \bibinfo {author} {\bibfnamefont {B.}~\bibnamefont {Gouraud}},
  \bibinfo {author} {\bibfnamefont {C.~T.}\ \bibnamefont {Ngai}}, \bibinfo
  {author} {\bibfnamefont {G.-L.}\ \bibnamefont {Schmid}}, \bibinfo {author}
  {\bibfnamefont {K.}~\bibnamefont {Hammerer}}, \ and\ \bibinfo {author}
  {\bibfnamefont {P.}~\bibnamefont {Treutlein}},\ }\bibfield  {title}
  {Light-mediated strong coupling between a mechanical oscillator and atomic
  spins 1 meter apart,\ }\href@noop {} {\bibfield  {journal} {\bibinfo
  {journal} {Science}\ }\textbf {\bibinfo {volume} {369}},\ \bibinfo {pages}
  {174} (\bibinfo {year} {2020})}\BibitemShut {NoStop}%
\bibitem [{\citenamefont {Van~Loo}\ \emph {et~al.}(2013)\citenamefont
  {Van~Loo}, \citenamefont {Fedorov}, \citenamefont {Lalumiere}, \citenamefont
  {Sanders}, \citenamefont {Blais},\ and\ \citenamefont
  {Wallraff}}]{van2013photon}%
  \BibitemOpen
  \bibfield  {author} {\bibinfo {author} {\bibfnamefont {A.~F.}\ \bibnamefont
  {Van~Loo}}, \bibinfo {author} {\bibfnamefont {A.}~\bibnamefont {Fedorov}},
  \bibinfo {author} {\bibfnamefont {K.}~\bibnamefont {Lalumiere}}, \bibinfo
  {author} {\bibfnamefont {B.~C.}\ \bibnamefont {Sanders}}, \bibinfo {author}
  {\bibfnamefont {A.}~\bibnamefont {Blais}}, \ and\ \bibinfo {author}
  {\bibfnamefont {A.}~\bibnamefont {Wallraff}},\ }\bibfield  {title}
  {Photon-mediated interactions between distant artificial atoms,\ }\href@noop
  {} {\bibfield  {journal} {\bibinfo  {journal} {Science}\ }\textbf {\bibinfo
  {volume} {342}},\ \bibinfo {pages} {1494} (\bibinfo {year}
  {2013})}\BibitemShut {NoStop}%
\bibitem [{\citenamefont {Hsu}\ \emph {et~al.}(2016)\citenamefont {Hsu},
  \citenamefont {Zhen}, \citenamefont {Stone}, \citenamefont {Joannopoulos},\
  and\ \citenamefont {Solja{\v{c}}i{\'c}}}]{hsu2016bound}%
  \BibitemOpen
  \bibfield  {author} {\bibinfo {author} {\bibfnamefont {C.~W.}\ \bibnamefont
  {Hsu}}, \bibinfo {author} {\bibfnamefont {B.}~\bibnamefont {Zhen}}, \bibinfo
  {author} {\bibfnamefont {A.~D.}\ \bibnamefont {Stone}}, \bibinfo {author}
  {\bibfnamefont {J.~D.}\ \bibnamefont {Joannopoulos}}, \ and\ \bibinfo
  {author} {\bibfnamefont {M.}~\bibnamefont {Solja{\v{c}}i{\'c}}},\ }\bibfield
  {title} {Bound states in the continuum,\ }\href@noop {} {\bibfield  {journal}
  {\bibinfo  {journal} {Nature Reviews Materials}\ }\textbf {\bibinfo {volume}
  {1}},\ \bibinfo {pages} {1} (\bibinfo {year} {2016})}\BibitemShut {NoStop}%
\bibitem [{\citenamefont {Rao}(2023)}]{SM}%
  \BibitemOpen
  \bibfield  {author} {\bibinfo {author} {\bibfnamefont {J.}~\bibnamefont
  {Rao}},\ }\bibfield  {title} {Supplementary material for "meter-scale strong
  coupling between magnons and photons",\ }\href@noop {} {\bibfield  {journal}
  {\bibinfo  {journal} {Supplementary material}\ } (\bibinfo {year}
  {2023})}\BibitemShut {NoStop}%
\bibitem [{\citenamefont {Rameshti}\ \emph {et~al.}(2022)\citenamefont
  {Rameshti}, \citenamefont {Kusminskiy}, \citenamefont {Haigh}, \citenamefont
  {Usami}, \citenamefont {Lachance-Quirion}, \citenamefont {Nakamura},
  \citenamefont {Hu}, \citenamefont {Tang}, \citenamefont {Bauer},\ and\
  \citenamefont {Blanter}}]{rameshti2022cavity}%
  \BibitemOpen
  \bibfield  {author} {\bibinfo {author} {\bibfnamefont {B.~Z.}\ \bibnamefont
  {Rameshti}}, \bibinfo {author} {\bibfnamefont {S.~V.}\ \bibnamefont
  {Kusminskiy}}, \bibinfo {author} {\bibfnamefont {J.~A.}\ \bibnamefont
  {Haigh}}, \bibinfo {author} {\bibfnamefont {K.}~\bibnamefont {Usami}},
  \bibinfo {author} {\bibfnamefont {D.}~\bibnamefont {Lachance-Quirion}},
  \bibinfo {author} {\bibfnamefont {Y.}~\bibnamefont {Nakamura}}, \bibinfo
  {author} {\bibfnamefont {C.-M.}\ \bibnamefont {Hu}}, \bibinfo {author}
  {\bibfnamefont {H.~X.}\ \bibnamefont {Tang}}, \bibinfo {author}
  {\bibfnamefont {G.~E.}\ \bibnamefont {Bauer}}, \ and\ \bibinfo {author}
  {\bibfnamefont {Y.~M.}\ \bibnamefont {Blanter}},\ }\bibfield  {title} {Cavity
  magnonics,\ }\href@noop {} {\bibfield  {journal} {\bibinfo  {journal}
  {Physics Reports}\ }\textbf {\bibinfo {volume} {979}},\ \bibinfo {pages} {1}
  (\bibinfo {year} {2022})}\BibitemShut {NoStop}%
\bibitem [{\citenamefont {Zhang}\ \emph {et~al.}(2014)\citenamefont {Zhang},
  \citenamefont {Zou}, \citenamefont {Jiang},\ and\ \citenamefont
  {Tang}}]{zhang2014strongly}%
  \BibitemOpen
  \bibfield  {author} {\bibinfo {author} {\bibfnamefont {X.}~\bibnamefont
  {Zhang}}, \bibinfo {author} {\bibfnamefont {C.-L.}\ \bibnamefont {Zou}},
  \bibinfo {author} {\bibfnamefont {L.}~\bibnamefont {Jiang}}, \ and\ \bibinfo
  {author} {\bibfnamefont {H.~X.}\ \bibnamefont {Tang}},\ }\bibfield  {title}
  {Strongly coupled magnons and cavity microwave photons,\ }\href@noop {}
  {\bibfield  {journal} {\bibinfo  {journal} {Physical Review Letters}\
  }\textbf {\bibinfo {volume} {113}},\ \bibinfo {pages} {156401} (\bibinfo
  {year} {2014})}\BibitemShut {NoStop}%
\bibitem [{\citenamefont {Huebl}\ \emph {et~al.}(2013)\citenamefont {Huebl},
  \citenamefont {Zollitsch}, \citenamefont {Lotze}, \citenamefont {Hocke},
  \citenamefont {Greifenstein}, \citenamefont {Marx}, \citenamefont {Gross},\
  and\ \citenamefont {Goennenwein}}]{huebl2013high}%
  \BibitemOpen
  \bibfield  {author} {\bibinfo {author} {\bibfnamefont {H.}~\bibnamefont
  {Huebl}}, \bibinfo {author} {\bibfnamefont {C.~W.}\ \bibnamefont
  {Zollitsch}}, \bibinfo {author} {\bibfnamefont {J.}~\bibnamefont {Lotze}},
  \bibinfo {author} {\bibfnamefont {F.}~\bibnamefont {Hocke}}, \bibinfo
  {author} {\bibfnamefont {M.}~\bibnamefont {Greifenstein}}, \bibinfo {author}
  {\bibfnamefont {A.}~\bibnamefont {Marx}}, \bibinfo {author} {\bibfnamefont
  {R.}~\bibnamefont {Gross}}, \ and\ \bibinfo {author} {\bibfnamefont {S.~T.}\
  \bibnamefont {Goennenwein}},\ }\bibfield  {title} {High cooperativity in
  coupled microwave resonator ferrimagnetic insulator hybrids,\ }\href@noop {}
  {\bibfield  {journal} {\bibinfo  {journal} {Physical Review Letters}\
  }\textbf {\bibinfo {volume} {111}},\ \bibinfo {pages} {127003} (\bibinfo
  {year} {2013})}\BibitemShut {NoStop}%
\bibitem [{\citenamefont {Goryachev}\ \emph {et~al.}(2014)\citenamefont
  {Goryachev}, \citenamefont {Farr}, \citenamefont {Creedon}, \citenamefont
  {Fan}, \citenamefont {Kostylev},\ and\ \citenamefont
  {Tobar}}]{goryachev2014high}%
  \BibitemOpen
  \bibfield  {author} {\bibinfo {author} {\bibfnamefont {M.}~\bibnamefont
  {Goryachev}}, \bibinfo {author} {\bibfnamefont {W.~G.}\ \bibnamefont {Farr}},
  \bibinfo {author} {\bibfnamefont {D.~L.}\ \bibnamefont {Creedon}}, \bibinfo
  {author} {\bibfnamefont {Y.}~\bibnamefont {Fan}}, \bibinfo {author}
  {\bibfnamefont {M.}~\bibnamefont {Kostylev}}, \ and\ \bibinfo {author}
  {\bibfnamefont {M.~E.}\ \bibnamefont {Tobar}},\ }\bibfield  {title}
  {High-cooperativity cavity qed with magnons at microwave frequencies,\
  }\href@noop {} {\bibfield  {journal} {\bibinfo  {journal} {Physical Review
  Applied}\ }\textbf {\bibinfo {volume} {2}},\ \bibinfo {pages} {054002}
  (\bibinfo {year} {2014})}\BibitemShut {NoStop}%
\bibitem [{\citenamefont {Zhang}\ \emph {et~al.}(2015)\citenamefont {Zhang},
  \citenamefont {Zou}, \citenamefont {Zhu}, \citenamefont {Marquardt},
  \citenamefont {Jiang},\ and\ \citenamefont {Tang}}]{zhang2015magnon}%
  \BibitemOpen
  \bibfield  {author} {\bibinfo {author} {\bibfnamefont {X.}~\bibnamefont
  {Zhang}}, \bibinfo {author} {\bibfnamefont {C.-L.}\ \bibnamefont {Zou}},
  \bibinfo {author} {\bibfnamefont {N.}~\bibnamefont {Zhu}}, \bibinfo {author}
  {\bibfnamefont {F.}~\bibnamefont {Marquardt}}, \bibinfo {author}
  {\bibfnamefont {L.}~\bibnamefont {Jiang}}, \ and\ \bibinfo {author}
  {\bibfnamefont {H.~X.}\ \bibnamefont {Tang}},\ }\bibfield  {title} {Magnon
  dark modes and gradient memory,\ }\href@noop {} {\bibfield  {journal}
  {\bibinfo  {journal} {Nature Communications}\ }\textbf {\bibinfo {volume}
  {6}},\ \bibinfo {pages} {8914} (\bibinfo {year} {2015})}\BibitemShut
  {NoStop}%
\bibitem [{\citenamefont {Wang}\ \emph {et~al.}(2019)\citenamefont {Wang},
  \citenamefont {Rao}, \citenamefont {Yang}, \citenamefont {Xu}, \citenamefont
  {Gui}, \citenamefont {Yao}, \citenamefont {You},\ and\ \citenamefont
  {Hu}}]{wang2019nonreciprocity}%
  \BibitemOpen
  \bibfield  {author} {\bibinfo {author} {\bibfnamefont {Y.-P.}\ \bibnamefont
  {Wang}}, \bibinfo {author} {\bibfnamefont {J.}~\bibnamefont {Rao}}, \bibinfo
  {author} {\bibfnamefont {Y.}~\bibnamefont {Yang}}, \bibinfo {author}
  {\bibfnamefont {P.-C.}\ \bibnamefont {Xu}}, \bibinfo {author} {\bibfnamefont
  {Y.}~\bibnamefont {Gui}}, \bibinfo {author} {\bibfnamefont {B.}~\bibnamefont
  {Yao}}, \bibinfo {author} {\bibfnamefont {J.}~\bibnamefont {You}}, \ and\
  \bibinfo {author} {\bibfnamefont {C.-M.}\ \bibnamefont {Hu}},\ }\bibfield
  {title} {Nonreciprocity and unidirectional invisibility in cavity magnonics,\
  }\href@noop {} {\bibfield  {journal} {\bibinfo  {journal} {Physical Review
  Letters}\ }\textbf {\bibinfo {volume} {123}},\ \bibinfo {pages} {127202}
  (\bibinfo {year} {2019})}\BibitemShut {NoStop}%
\bibitem [{\citenamefont {Wang}\ \emph {et~al.}(2021)\citenamefont {Wang},
  \citenamefont {Yuan}, \citenamefont {Cao}, \citenamefont {Li}, \citenamefont
  {Duine},\ and\ \citenamefont {Yan}}]{wang2021magnonic}%
  \BibitemOpen
  \bibfield  {author} {\bibinfo {author} {\bibfnamefont {Z.}~\bibnamefont
  {Wang}}, \bibinfo {author} {\bibfnamefont {H.}~\bibnamefont {Yuan}}, \bibinfo
  {author} {\bibfnamefont {Y.}~\bibnamefont {Cao}}, \bibinfo {author}
  {\bibfnamefont {Z.-X.}\ \bibnamefont {Li}}, \bibinfo {author} {\bibfnamefont
  {R.~A.}\ \bibnamefont {Duine}}, \ and\ \bibinfo {author} {\bibfnamefont
  {P.}~\bibnamefont {Yan}},\ }\bibfield  {title} {Magnonic frequency comb
  through nonlinear magnon-skyrmion scattering,\ }\href@noop {} {\bibfield
  {journal} {\bibinfo  {journal} {Physical Review Letters}\ }\textbf {\bibinfo
  {volume} {127}},\ \bibinfo {pages} {037202} (\bibinfo {year}
  {2021})}\BibitemShut {NoStop}%
\bibitem [{\citenamefont {Yao}\ \emph {et~al.}(2023)\citenamefont {Yao},
  \citenamefont {Gui}, \citenamefont {Rao}, \citenamefont {Zhang},
  \citenamefont {Lu},\ and\ \citenamefont {Hu}}]{PhysRevLett.130.146702}%
  \BibitemOpen
  \bibfield  {author} {\bibinfo {author} {\bibfnamefont {B.}~\bibnamefont
  {Yao}}, \bibinfo {author} {\bibfnamefont {Y.~S.}\ \bibnamefont {Gui}},
  \bibinfo {author} {\bibfnamefont {J.~W.}\ \bibnamefont {Rao}}, \bibinfo
  {author} {\bibfnamefont {Y.~H.}\ \bibnamefont {Zhang}}, \bibinfo {author}
  {\bibfnamefont {W.}~\bibnamefont {Lu}}, \ and\ \bibinfo {author}
  {\bibfnamefont {C.-M.}\ \bibnamefont {Hu}},\ }\bibfield  {title} {Coherent
  microwave emission of gain-driven polaritons,\ }\href {\doibase
  10.1103/PhysRevLett.130.146702} {\bibfield  {journal} {\bibinfo  {journal}
  {Phys. Rev. Lett.}\ }\textbf {\bibinfo {volume} {130}},\ \bibinfo {pages}
  {146702} (\bibinfo {year} {2023})}\BibitemShut {NoStop}%
\bibitem [{\citenamefont {Van~der Pol}(1926)}]{van1926}%
  \BibitemOpen
  \bibfield  {author} {\bibinfo {author} {\bibfnamefont {B.}~\bibnamefont
  {Van~der Pol}},\ }\bibfield  {title} {On relaxiation oscillations,\
  }\href@noop {} {\bibfield  {journal} {\bibinfo  {journal} {The London,
  Edinburgh, and Dublin Philosophical Magazine and Journal of Science}\
  }\textbf {\bibinfo {volume} {2}},\ \bibinfo {pages} {978} (\bibinfo {year}
  {1926})}\BibitemShut {NoStop}%
\bibitem [{\citenamefont {Dutta}\ and\ \citenamefont
  {Cooper}(2019)}]{dutta2019critical}%
  \BibitemOpen
  \bibfield  {author} {\bibinfo {author} {\bibfnamefont {S.}~\bibnamefont
  {Dutta}}\ and\ \bibinfo {author} {\bibfnamefont {N.~R.}\ \bibnamefont
  {Cooper}},\ }\bibfield  {title} {Critical response of a quantum van der pol
  oscillator,\ }\href@noop {} {\bibfield  {journal} {\bibinfo  {journal}
  {Physical Review Letters}\ }\textbf {\bibinfo {volume} {123}},\ \bibinfo
  {pages} {250401} (\bibinfo {year} {2019})}\BibitemShut {NoStop}%
\bibitem [{\citenamefont {L{\"o}rch}\ \emph {et~al.}(2016)\citenamefont
  {L{\"o}rch}, \citenamefont {Amitai}, \citenamefont {Nunnenkamp},\ and\
  \citenamefont {Bruder}}]{lorch2016genuine}%
  \BibitemOpen
  \bibfield  {author} {\bibinfo {author} {\bibfnamefont {N.}~\bibnamefont
  {L{\"o}rch}}, \bibinfo {author} {\bibfnamefont {E.}~\bibnamefont {Amitai}},
  \bibinfo {author} {\bibfnamefont {A.}~\bibnamefont {Nunnenkamp}}, \ and\
  \bibinfo {author} {\bibfnamefont {C.}~\bibnamefont {Bruder}},\ }\bibfield
  {title} {Genuine quantum signatures in synchronization of anharmonic
  self-oscillators,\ }\href@noop {} {\bibfield  {journal} {\bibinfo  {journal}
  {Physical Review Letters}\ }\textbf {\bibinfo {volume} {117}},\ \bibinfo
  {pages} {073601} (\bibinfo {year} {2016})}\BibitemShut {NoStop}%
\bibitem [{\citenamefont {Van Der~Pol}\ and\ \citenamefont {Van
  Der~Mark}(1928)}]{van1928lxxii}%
  \BibitemOpen
  \bibfield  {author} {\bibinfo {author} {\bibfnamefont {B.}~\bibnamefont {Van
  Der~Pol}}\ and\ \bibinfo {author} {\bibfnamefont {J.}~\bibnamefont {Van
  Der~Mark}},\ }\bibfield  {title} {Lxxii. the heartbeat considered as a
  relaxation oscillation, and an electrical model of the heart,\ }\href@noop {}
  {\bibfield  {journal} {\bibinfo  {journal} {The London, Edinburgh, and Dublin
  Philosophical Magazine and Journal of Science}\ }\textbf {\bibinfo {volume}
  {6}},\ \bibinfo {pages} {763} (\bibinfo {year} {1928})}\BibitemShut {NoStop}%
\bibitem [{\citenamefont {Uemukai}\ \emph {et~al.}(2000)\citenamefont
  {Uemukai}, \citenamefont {Suhara}, \citenamefont {Yutani}, \citenamefont
  {Shimada}, \citenamefont {Fukumoto}, \citenamefont {Nishihara},\ and\
  \citenamefont {Larsson}}]{uemukai2000tunable}%
  \BibitemOpen
  \bibfield  {author} {\bibinfo {author} {\bibfnamefont {M.}~\bibnamefont
  {Uemukai}}, \bibinfo {author} {\bibfnamefont {T.}~\bibnamefont {Suhara}},
  \bibinfo {author} {\bibfnamefont {K.}~\bibnamefont {Yutani}}, \bibinfo
  {author} {\bibfnamefont {N.}~\bibnamefont {Shimada}}, \bibinfo {author}
  {\bibfnamefont {Y.}~\bibnamefont {Fukumoto}}, \bibinfo {author}
  {\bibfnamefont {H.}~\bibnamefont {Nishihara}}, \ and\ \bibinfo {author}
  {\bibfnamefont {A.}~\bibnamefont {Larsson}},\ }\bibfield  {title} {Tunable
  external-cavity semiconductor laser using monolithically integrated tapered
  amplifier and grating coupler for collimation,\ }\href@noop {} {\bibfield
  {journal} {\bibinfo  {journal} {IEEE Photonics Technology Letters}\ }\textbf
  {\bibinfo {volume} {12}},\ \bibinfo {pages} {1607} (\bibinfo {year}
  {2000})}\BibitemShut {NoStop}%
\bibitem [{\citenamefont {Leibfried}\ \emph {et~al.}(2003)\citenamefont
  {Leibfried}, \citenamefont {Blatt}, \citenamefont {Monroe},\ and\
  \citenamefont {Wineland}}]{leibfried2003quantum}%
  \BibitemOpen
  \bibfield  {author} {\bibinfo {author} {\bibfnamefont {D.}~\bibnamefont
  {Leibfried}}, \bibinfo {author} {\bibfnamefont {R.}~\bibnamefont {Blatt}},
  \bibinfo {author} {\bibfnamefont {C.}~\bibnamefont {Monroe}}, \ and\ \bibinfo
  {author} {\bibfnamefont {D.}~\bibnamefont {Wineland}},\ }\bibfield  {title}
  {Quantum dynamics of single trapped ions,\ }\href@noop {} {\bibfield
  {journal} {\bibinfo  {journal} {Reviews of Modern Physics}\ }\textbf
  {\bibinfo {volume} {75}},\ \bibinfo {pages} {281} (\bibinfo {year}
  {2003})}\BibitemShut {NoStop}%
\bibitem [{\citenamefont {Gilles}\ and\ \citenamefont
  {Knight}(1993)}]{gilles1993two}%
  \BibitemOpen
  \bibfield  {author} {\bibinfo {author} {\bibfnamefont {L.}~\bibnamefont
  {Gilles}}\ and\ \bibinfo {author} {\bibfnamefont {P.}~\bibnamefont
  {Knight}},\ }\bibfield  {title} {Two-photon absorption and nonclassical
  states of light,\ }\href@noop {} {\bibfield  {journal} {\bibinfo  {journal}
  {Physical Review A}\ }\textbf {\bibinfo {volume} {48}},\ \bibinfo {pages}
  {1582} (\bibinfo {year} {1993})}\BibitemShut {NoStop}%
\bibitem [{\citenamefont {Lee}\ and\ \citenamefont
  {Sadeghpour}(2013)}]{lee2013quantum}%
  \BibitemOpen
  \bibfield  {author} {\bibinfo {author} {\bibfnamefont {T.~E.}\ \bibnamefont
  {Lee}}\ and\ \bibinfo {author} {\bibfnamefont {H.}~\bibnamefont
  {Sadeghpour}},\ }\bibfield  {title} {Quantum synchronization of quantum van
  der pol oscillators with trapped ions,\ }\href@noop {} {\bibfield  {journal}
  {\bibinfo  {journal} {Physical Review Letters}\ }\textbf {\bibinfo {volume}
  {111}},\ \bibinfo {pages} {234101} (\bibinfo {year} {2013})}\BibitemShut
  {NoStop}%
\bibitem [{\citenamefont {Rao}\ \emph {et~al.}(2020)\citenamefont {Rao},
  \citenamefont {Wang}, \citenamefont {Yang}, \citenamefont {Yu}, \citenamefont
  {Gui}, \citenamefont {Fan}, \citenamefont {Xue},\ and\ \citenamefont
  {Hu}}]{rao2020interactions}%
  \BibitemOpen
  \bibfield  {author} {\bibinfo {author} {\bibfnamefont {J.}~\bibnamefont
  {Rao}}, \bibinfo {author} {\bibfnamefont {Y.}~\bibnamefont {Wang}}, \bibinfo
  {author} {\bibfnamefont {Y.}~\bibnamefont {Yang}}, \bibinfo {author}
  {\bibfnamefont {T.}~\bibnamefont {Yu}}, \bibinfo {author} {\bibfnamefont
  {Y.}~\bibnamefont {Gui}}, \bibinfo {author} {\bibfnamefont {X.}~\bibnamefont
  {Fan}}, \bibinfo {author} {\bibfnamefont {D.}~\bibnamefont {Xue}}, \ and\
  \bibinfo {author} {\bibfnamefont {C.-M.}\ \bibnamefont {Hu}},\ }\bibfield
  {title} {Interactions between a magnon mode and a cavity photon mode mediated
  by traveling photons,\ }\href@noop {} {\bibfield  {journal} {\bibinfo
  {journal} {Physical Review B}\ }\textbf {\bibinfo {volume} {101}},\ \bibinfo
  {pages} {064404} (\bibinfo {year} {2020})}\BibitemShut {NoStop}%
\bibitem [{\citenamefont {Yang}\ \emph {et~al.}(2018)\citenamefont {Yang},
  \citenamefont {Harder}, \citenamefont {Rao}, \citenamefont {Yao},
  \citenamefont {Lu}, \citenamefont {Gui},\ and\ \citenamefont
  {Hu}}]{yang2018influence}%
  \BibitemOpen
  \bibfield  {author} {\bibinfo {author} {\bibfnamefont {Y.}~\bibnamefont
  {Yang}}, \bibinfo {author} {\bibfnamefont {M.}~\bibnamefont {Harder}},
  \bibinfo {author} {\bibfnamefont {J.}~\bibnamefont {Rao}}, \bibinfo {author}
  {\bibfnamefont {B.}~\bibnamefont {Yao}}, \bibinfo {author} {\bibfnamefont
  {W.}~\bibnamefont {Lu}}, \bibinfo {author} {\bibfnamefont {Y.}~\bibnamefont
  {Gui}}, \ and\ \bibinfo {author} {\bibfnamefont {C.-M.}\ \bibnamefont {Hu}},\
  }\bibfield  {title} {Influence of stripline coupling on the magnetostatic
  mode line width of an yttrium-iron-garnet sphere,\ }\href@noop {} {\bibfield
  {journal} {\bibinfo  {journal} {AIP Advances}\ }\textbf {\bibinfo {volume}
  {8}},\ \bibinfo {pages} {075315} (\bibinfo {year} {2018})}\BibitemShut
  {NoStop}%
\bibitem [{\citenamefont {Li}\ \emph {et~al.}(2019)\citenamefont {Li},
  \citenamefont {Naletov}, \citenamefont {Klein}, \citenamefont {Prieto},
  \citenamefont {Mu{\~n}oz}, \citenamefont {Cros}, \citenamefont {Bortolotti},
  \citenamefont {Anane}, \citenamefont {Serpico},\ and\ \citenamefont
  {De~Loubens}}]{li2019nutation}%
  \BibitemOpen
  \bibfield  {author} {\bibinfo {author} {\bibfnamefont {Y.}~\bibnamefont
  {Li}}, \bibinfo {author} {\bibfnamefont {V.~V.}\ \bibnamefont {Naletov}},
  \bibinfo {author} {\bibfnamefont {O.}~\bibnamefont {Klein}}, \bibinfo
  {author} {\bibfnamefont {J.~L.}\ \bibnamefont {Prieto}}, \bibinfo {author}
  {\bibfnamefont {M.}~\bibnamefont {Mu{\~n}oz}}, \bibinfo {author}
  {\bibfnamefont {V.}~\bibnamefont {Cros}}, \bibinfo {author} {\bibfnamefont
  {P.}~\bibnamefont {Bortolotti}}, \bibinfo {author} {\bibfnamefont
  {A.}~\bibnamefont {Anane}}, \bibinfo {author} {\bibfnamefont
  {C.}~\bibnamefont {Serpico}}, \ and\ \bibinfo {author} {\bibfnamefont
  {G.}~\bibnamefont {De~Loubens}},\ }\bibfield  {title} {Nutation spectroscopy
  of a nanomagnet driven into deeply nonlinear ferromagnetic resonance,\
  }\href@noop {} {\bibfield  {journal} {\bibinfo  {journal} {Physical Review
  X}\ }\textbf {\bibinfo {volume} {9}},\ \bibinfo {pages} {041036} (\bibinfo
  {year} {2019})}\BibitemShut {NoStop}%
\bibitem [{\citenamefont {Pippard}(2007)}]{pippard2007physics}%
  \BibitemOpen
  \bibfield  {author} {\bibinfo {author} {\bibfnamefont {A.~B.}\ \bibnamefont
  {Pippard}},\ }\bibfield  {title} {The physics of vibration,\ }\href@noop {}
  {\bibfield  {journal} {\bibinfo  {journal} {The Physics of Vibration}\ }
  (\bibinfo {year} {2007})}\BibitemShut {NoStop}%
\bibitem [{\citenamefont {Rao}\ \emph {et~al.}(2021)\citenamefont {Rao},
  \citenamefont {Xu}, \citenamefont {Gui}, \citenamefont {Wang}, \citenamefont
  {Yang}, \citenamefont {Yao}, \citenamefont {Dietrich}, \citenamefont
  {Bridges}, \citenamefont {Fan}, \citenamefont {Xue} \emph
  {et~al.}}]{rao2021interferometric}%
  \BibitemOpen
  \bibfield  {author} {\bibinfo {author} {\bibfnamefont {J.}~\bibnamefont
  {Rao}}, \bibinfo {author} {\bibfnamefont {P.}~\bibnamefont {Xu}}, \bibinfo
  {author} {\bibfnamefont {Y.}~\bibnamefont {Gui}}, \bibinfo {author}
  {\bibfnamefont {Y.}~\bibnamefont {Wang}}, \bibinfo {author} {\bibfnamefont
  {Y.}~\bibnamefont {Yang}}, \bibinfo {author} {\bibfnamefont {B.}~\bibnamefont
  {Yao}}, \bibinfo {author} {\bibfnamefont {J.}~\bibnamefont {Dietrich}},
  \bibinfo {author} {\bibfnamefont {G.}~\bibnamefont {Bridges}}, \bibinfo
  {author} {\bibfnamefont {X.}~\bibnamefont {Fan}}, \bibinfo {author}
  {\bibfnamefont {D.}~\bibnamefont {Xue}},  \emph {et~al.},\ }\bibfield
  {title} {Interferometric control of magnon-induced nearly perfect absorption
  in cavity magnonics,\ }\href@noop {} {\bibfield  {journal} {\bibinfo
  {journal} {Nature Communications}\ }\textbf {\bibinfo {volume} {12}},\
  \bibinfo {pages} {1} (\bibinfo {year} {2021})}\BibitemShut {NoStop}%
\bibitem [{\citenamefont {Grigoryan}\ \emph {et~al.}(2018)\citenamefont
  {Grigoryan}, \citenamefont {Shen},\ and\ \citenamefont
  {Xia}}]{grigoryan2018synchronized}%
  \BibitemOpen
  \bibfield  {author} {\bibinfo {author} {\bibfnamefont {V.~L.}\ \bibnamefont
  {Grigoryan}}, \bibinfo {author} {\bibfnamefont {K.}~\bibnamefont {Shen}}, \
  and\ \bibinfo {author} {\bibfnamefont {K.}~\bibnamefont {Xia}},\ }\bibfield
  {title} {Synchronized spin-photon coupling in a microwave cavity,\
  }\href@noop {} {\bibfield  {journal} {\bibinfo  {journal} {Physical Review
  B}\ }\textbf {\bibinfo {volume} {98}},\ \bibinfo {pages} {024406} (\bibinfo
  {year} {2018})}\BibitemShut {NoStop}%
\bibitem [{\citenamefont {Yu}\ \emph {et~al.}(2019)\citenamefont {Yu},
  \citenamefont {Wang}, \citenamefont {Yuan}, \citenamefont {Xiao} }]{yu2019prediction}%
  \BibitemOpen
  \bibfield  {author} {\bibinfo {author} {\bibfnamefont {W.}~\bibnamefont
  {Yu}}, \bibinfo {author} {\bibfnamefont {J.}~\bibnamefont {Wang}}, \bibinfo
  {author} {\bibfnamefont {H.}~\bibnamefont {Yuan}}, \bibinfo {author}
  {\bibfnamefont {J.}~\bibnamefont {Xiao}},\ }\bibfield
  {title} {Prediction of attractive level crossing via a dissipative mode,\
  }\href@noop {} {\bibfield  {journal} {\bibinfo  {journal} {Physical Review
  Letters}\ }\textbf {\bibinfo {volume} {123}},\ \bibinfo {pages} {227201}
  (\bibinfo {year} {2019})}\BibitemShut {NoStop}%
\bibitem [{\citenamefont {Bhoi}\ \emph {et~al.}(2019)\citenamefont {Bhoi},
  \citenamefont {Kim}, \citenamefont {Jang}, \citenamefont {Kim}, \citenamefont
  {Yang}, \citenamefont {Cho},\ and\ \citenamefont {Kim}}]{bhoi2019abnormal}%
  \BibitemOpen
  \bibfield  {author} {\bibinfo {author} {\bibfnamefont {B.}~\bibnamefont
  {Bhoi}}, \bibinfo {author} {\bibfnamefont {B.}~\bibnamefont {Kim}}, \bibinfo
  {author} {\bibfnamefont {S.-H.}\ \bibnamefont {Jang}}, \bibinfo {author}
  {\bibfnamefont {J.}~\bibnamefont {Kim}}, \bibinfo {author} {\bibfnamefont
  {J.}~\bibnamefont {Yang}}, \bibinfo {author} {\bibfnamefont {Y.-J.}\
  \bibnamefont {Cho}}, \ and\ \bibinfo {author} {\bibfnamefont {S.-K.}\
  \bibnamefont {Kim}},\ }\bibfield  {title} {Abnormal anticrossing effect in
  photon-magnon coupling,\ }\href@noop {} {\bibfield  {journal} {\bibinfo
  {journal} {Physical Review B}\ }\textbf {\bibinfo {volume} {99}},\ \bibinfo
  {pages} {134426} (\bibinfo {year} {2019})}\BibitemShut {NoStop}%
\bibitem [{\citenamefont {Boventer}\ \emph {et~al.}(2019)\citenamefont
  {Boventer}, \citenamefont {Kl{\"a}ui}, \citenamefont {Mac{\^e}do},\ and\
  \citenamefont {Weides}}]{boventer2019steering}%
  \BibitemOpen
  \bibfield  {author} {\bibinfo {author} {\bibfnamefont {I.}~\bibnamefont
  {Boventer}}, \bibinfo {author} {\bibfnamefont {M.}~\bibnamefont {Kl{\"a}ui}},
  \bibinfo {author} {\bibfnamefont {R.}~\bibnamefont {Mac{\^e}do}}, \ and\
  \bibinfo {author} {\bibfnamefont {M.}~\bibnamefont {Weides}},\ }\bibfield
  {title} {Steering between level repulsion and attraction: broad tunability of
  two-port driven cavity magnon-polaritons,\ }\href@noop {} {\bibfield
  {journal} {\bibinfo  {journal} {New Journal of Physics}\ }\textbf {\bibinfo
  {volume} {21}},\ \bibinfo {pages} {125001} (\bibinfo {year}
  {2019})}\BibitemShut {NoStop}%
\bibitem [{\citenamefont {Harder}\ \emph {et~al.}(2018)\citenamefont {Harder},
  \citenamefont {Yang}, \citenamefont {Yao}, \citenamefont {Yu}, \citenamefont
  {Rao}, \citenamefont {Gui}, \citenamefont {Stamps},\ and\ \citenamefont
  {Hu}}]{harder2018level}%
  \BibitemOpen
  \bibfield  {author} {\bibinfo {author} {\bibfnamefont {M.}~\bibnamefont
  {Harder}}, \bibinfo {author} {\bibfnamefont {Y.}~\bibnamefont {Yang}},
  \bibinfo {author} {\bibfnamefont {B.}~\bibnamefont {Yao}}, \bibinfo {author}
  {\bibfnamefont {C.}~\bibnamefont {Yu}}, \bibinfo {author} {\bibfnamefont
  {J.}~\bibnamefont {Rao}}, \bibinfo {author} {\bibfnamefont {Y.}~\bibnamefont
  {Gui}}, \bibinfo {author} {\bibfnamefont {R.}~\bibnamefont {Stamps}}, \ and\
  \bibinfo {author} {\bibfnamefont {C.-M.}\ \bibnamefont {Hu}},\ }\bibfield
  {title} {Level attraction due to dissipative magnon-photon coupling,\
  }\href@noop {} {\bibfield  {journal} {\bibinfo  {journal} {Physical Review
  Letters}\ }\textbf {\bibinfo {volume} {121}},\ \bibinfo {pages} {137203}
  (\bibinfo {year} {2018})}\BibitemShut {NoStop}%
\bibitem [{\citenamefont {Rao}\ \emph {et~al.}(2019)\citenamefont {Rao},
  \citenamefont {Kaur}, \citenamefont {Yao}, \citenamefont {Edwards},
  \citenamefont {Zhao}, \citenamefont {Fan}, \citenamefont {Xue}, \citenamefont
  {Silva}, \citenamefont {Gui},\ and\ \citenamefont {Hu}}]{rao2019analogue}%
  \BibitemOpen
  \bibfield  {author} {\bibinfo {author} {\bibfnamefont {J.}~\bibnamefont
  {Rao}}, \bibinfo {author} {\bibfnamefont {S.}~\bibnamefont {Kaur}}, \bibinfo
  {author} {\bibfnamefont {B.}~\bibnamefont {Yao}}, \bibinfo {author}
  {\bibfnamefont {E.}~\bibnamefont {Edwards}}, \bibinfo {author} {\bibfnamefont
  {Y.}~\bibnamefont {Zhao}}, \bibinfo {author} {\bibfnamefont {X.}~\bibnamefont
  {Fan}}, \bibinfo {author} {\bibfnamefont {D.}~\bibnamefont {Xue}}, \bibinfo
  {author} {\bibfnamefont {T.~J.}\ \bibnamefont {Silva}}, \bibinfo {author}
  {\bibfnamefont {Y.}~\bibnamefont {Gui}}, \ and\ \bibinfo {author}
  {\bibfnamefont {C.-M.}\ \bibnamefont {Hu}},\ }\bibfield  {title} {Analogue of
  dynamic hall effect in cavity magnon polariton system and coherently
  controlled logic device,\ }\href@noop {} {\bibfield  {journal} {\bibinfo
  {journal} {Nature Communications}\ }\textbf {\bibinfo {volume} {10}},\
  \bibinfo {pages} {2934} (\bibinfo {year} {2019})}\BibitemShut {NoStop}%
\bibitem [{\citenamefont {Yu}\ \emph {et~al.}(2020)\citenamefont {Yu},
  \citenamefont {Zhang}, \citenamefont {Sharma}, \citenamefont {Zhang},
  \citenamefont {Blanter},\ and\ \citenamefont {Bauer}}]{yu2020magnon}%
  \BibitemOpen
  \bibfield  {author} {\bibinfo {author} {\bibfnamefont {T.}~\bibnamefont
  {Yu}}, \bibinfo {author} {\bibfnamefont {Y.-X.}\ \bibnamefont {Zhang}},
  \bibinfo {author} {\bibfnamefont {S.}~\bibnamefont {Sharma}}, \bibinfo
  {author} {\bibfnamefont {X.}~\bibnamefont {Zhang}}, \bibinfo {author}
  {\bibfnamefont {Y.~M.}\ \bibnamefont {Blanter}}, \ and\ \bibinfo {author}
  {\bibfnamefont {G.~E.}\ \bibnamefont {Bauer}},\ }\bibfield  {title} {Magnon
  accumulation in chirally coupled magnets,\ }\href@noop {} {\bibfield
  {journal} {\bibinfo  {journal} {Physical Review Letters}\ }\textbf {\bibinfo
  {volume} {124}},\ \bibinfo {pages} {107202} (\bibinfo {year}
  {2020})}\BibitemShut {NoStop}%
\end{thebibliography}
\end{document}